\documentclass[10pt]{article}
\usepackage{graphicx}
\usepackage{amssymb}
\usepackage{amsmath}
\usepackage{bm}
\usepackage{upmath}
\usepackage{color}
\usepackage{url}

\newcommand{\be}{\begin{equation}}
\newcommand{\ee}{\end{equation}}
\newcommand{\ba}{\begin{eqnarray}}
\newcommand{\ea}{\end{eqnarray}}
\renewcommand{\l}{\label}
\newcommand{\f}{\frac}

\renewcommand{\a}{\alpha}
\renewcommand{\b}{\beta}
\renewcommand{\d}{\delta}

\newcommand{\p}{\partial}
\renewcommand{\le}{\left}
\renewcommand{\r}{\right}

\newcommand{\aj}{{Astron. J. (USA)}}

\newcommand{\pla}{{{Phys. Lett.}  A}}
\newcommand{\plb}{{{Phys. Lett.}  B}}

\newcommand{\prd}{{{Phys. Rev.} D}}

\newcommand{\prep}{{Phys. Reports}}

\newcommand{\aap}{{Astron. Astrophys.}}

\def\xe{({\bm x}_E)}

\begin{document}

\setcounter{figure}{0}
\setcounter{table}{0}
\setcounter{footnote}{0}
\setcounter{equation}{0}

\vspace*{0.5cm}

\noindent\centerline{\Large\bf AN EXTENSION OF THE IAU FRAMEWORK}
\centerline{\Large\bf FOR REFERENCE SYSTEMS}
\vspace*{0.7cm}

\noindent\hspace*{1.5cm} S.M. KOPEIKIN\\
\noindent\hspace*{1.5cm} Department of Physics \& Astronomy, University of Missouri-Columbia\\
\noindent\hspace*{1.5cm} 223 Physics Bldg., Columbia, Missouri 65211, USA\\
\noindent\hspace*{1.5cm} e-mail: kopeikins@missouri.edu\\

\vspace*{0.5cm}

\noindent {\large\bf ABSTRACT.}

IAU 2000 resolutions on the reference frames set up a solid theoretical foundation for implementing general relativity in astronomical data processing algorithms and for unambiguous interpretation of measured relativistic effects. We discuss possible directions for further theoretical development of the IAU resolutions aimed to take into account the decadal progress in observational techniques and computer-based technologies. We address the following subjects: 1) space-time transformations and the structure of the metric tensor; -2) PPN parameters and gauge invariance of equations of motion; -3) astronomical reference frames for cosmological applications.

\section{\large INTRODUCTION}

Experimental exploration of the nature of space-time demands establishment of a common theoretical platform linking a theory of gravitational field to astronomical observations. This platform should be build on the basis of a complete theory of gravity like general theory of relativity, that describes both the properties of space-time, gravitational field and observables. New generation of microarcsecond astrometry satellites like SIM Lite \footnote{The Astro2010 Decadal Survey (available at \url{http://sites.nationalacademies.org/bpa/BPA_049810}) did not recommend SIM Lite for development this decade.}
and a cornerstone mission of ESA - Gaia, require such a novel approach for an unambiguous
interpretation of astrometric data obtained from the on-board
optical instruments. Advanced inertial reference frame is required for unambiguous physical interpretation of gravitomagnetic precession of LAGEOS satellite and LLR observations \cite{ciuf}. Recent breakthroughs in technology of drag-free satellites, clocks, lasers, optical and radio interferometers and new demands of experimental gravitational physics \cite{leh,hlnt} make it necessary to incorporate the parameterized post-Newtonian formalism \cite{will} to the procedure of construction of relativistic local frames around Earth and other bodies of the solar system \cite{kv,xie}.
The domain of applicability of the IAU relativistic theory of
reference frames \cite{iau2} is to be also extended outside the solar system \cite{kopg} to take into account the impact of the Hubble expansion on the solutions of the gravity field equations and the equations of motion of the bodies.

In what follows, Latin indices takes values 1,2,3; the Greek indices run from 0 to 3. Repeated indices imply the Einstein summation rule. The unit matrix $\delta_{ij}={\rm diag}(1,1,1)$ and the fully anti-symmetric symbol $\epsilon_{ijk}$ is subject to $\epsilon_{123}=1$. The Minkowski metric $\eta_{\a\b}={\rm diag}(-1,1,1,1)$. Greek indices are raised and lowered with the Minkowski metric, Latin indices are raised and lowered with the unit matrix. Bold italic letters ${\bm a}$, ${\bm b}$, etc., denote spatial vectors. A dot and a cross between two spatial vectors denote the Euclidean scalar and vector products respectively. Partial derivative with respect to spatial coordinates $x^i$ are denoted as $\p/\p x^i$ or ${\vec\nabla}$.

\section{\large STANDARD IAU FRAMEWORK}
The IAU resolutions are based on the first post-Newtonian
approximation of general relativity which is a conceptual basis of the fundamental astronomy in the solar system \cite{iau06}.
Barycentric Celestial Reference System (BCRS), $x^\a=(ct,{\bm x})$,
is defined in terms of a metric tensor $g_{\a\b}$ with components
\begin{eqnarray}
g_{00} &=& - 1 + \frac{2 w}{c^2} - \frac{2w^2}{c^4} + O(c^{-5})\;,
\label{5} \\
g_{0i} &=& - \frac{4 w^i}{c^3} + O(c^{-5})\;,\label{6}
 \\
g_{ij} &=& \delta_{ij}
\left( 1 + \frac{2w}{c^2} \right) + O(c^{-4})\; .
\label{7}
\end{eqnarray}
Here, the post-Newtonian gravitational potentials $w$ and $w^i$ are defined by solving the
gravity field equations
\begin{eqnarray}
\label{eq1}
\Box w&=&-4\pi G\sigma\;,\\
\label{eq2}
\Box w^i&=&-4\pi G\sigma^i\;,
\end{eqnarray}
where $\Box\equiv -c^{-2}\partial^2/\partial t^2+\nabla^2$ is the wave operator,
$\sigma = c^{-2}(T^{00} + T^{ss}),$ $
\sigma^i = c^{-1}T^{0i}$, and
$T^{\mu\nu}$ are the components of the stress-energy tensor of the solar system bodies, $T^{ss}= T^{11} + T^{22} + T^{33}$.

Equations (\ref{eq1}), (\ref{eq2})
are solved by iterations
\begin{eqnarray}
w(t,{\bm x}) &=& G \int
\frac{\sigma(t, {\bm x}')d^3 x'}{\vert {\bm x} - {\bm x}' \vert}
 + \frac{G }{2c^2}  \frac{\partial^2}{\partial t^2}
\int d^3 x'  \sigma(t,{\bm x}') \vert {\bm x} - {\bm x}' \vert +O(c^{-4})\; ,
\label{8}\\\label{9}
w^i(t,{\bm x}) &=& G \int\frac{\sigma^i (t,{\bm x}') d^3 x'}{
\vert{\bm x} - {\bm x}' \vert } +O(c^{-2})\;,
\end{eqnarray}
which are to be substituted to the metric tensor (\ref{5})--(\ref{7}). Each of the potentials, $w$ and $w^i$, can be linearly decomposed in two pieces
\ba\l{eq3}
w&=&w_E+{\bar w}\;,\\\l{eq4}
w^i&=&w_E^i+{\bar w}^i\;,
\ea
where $w_E$ and $w_E^i$ are BCRS potentials depending on the distribution of mass and current only inside the Earth, and  ${\bar w}_E$ and ${\bar w}_E^i$ are gravitational potentials of external bodies.

Geocentric Celestial Reference System (GCRS) is denoted $X^\a=(cT, {\bm X})$.
It has the metric tensor $G_{\a\b}$ with components
\begin{eqnarray}
G_{00} &=& - 1 + \frac{2 W}{c^2} - \frac{2W^2 }{ c^4} + O(c^{-5})\;,
\label{11} \\\label{12}
G_{0i} &=& - \frac{4 W^i }{c^3} + O(c^{-5})\;,
 \\
G_{ij} &=& \delta_{ij}
\left( 1 + \frac{2 W }{ c^2} \right) + O(c^{-4})\; .
\label{13}
\end{eqnarray}
Here $W = W(T,{\bm X})$ is the post-Newtonian gravitational potential and $W^i(T,{\bm X})$ is a vector-potential both expressed in the geocentric coordinates. They satisfy to the same type of the wave equations (\ref{eq1}), (\ref{eq2}). Planetocentric metric for any planet can be introduced in the same way as the GCRS.

The geocentric potentials, $W_E$ and $W_E^i$, are split into three parts
\begin{eqnarray}
W(T,{\bm X}) &=& W_E(T,{\bm X})
+ W_{\rm kin}(T,{\bm X})+W_{\rm dyn}(T,{\bm X})
\;,
\label{14}\\
\label{15}
W^i(T,{\bm X}) &=& W^i_E(T,{\bm X})
+ W^i_{\rm kin}(T,{\bm X})+W^i_{\rm dyn}(T,{\bm X})
\;.
\end{eqnarray}
associated respectively with the gravitational field of the Earth, external tidal field and kinematic inertial force. IAU resolutions imply that the external and kinematic parts must vanish at the
geocenter and admit an expansion in powers of ${\bm X}$ \cite{iau2}. Geopotentials  $W_E$ and $W^i_E$
are defined in the same way as $w_E$ and $w_E^i$ in equations (\ref{8})-(\ref{9}) but with $\sigma$ and $\sigma^i$
calculated in the GCRS. They are related to the barycentric gravitational potentials $w_E$ and
$w^i_E$ by the post-Newtonian transformations \cite{bk,iau2}.

The kinematic contributions are linear in the GCRS spatial coordinates ${\bm X}$
\begin{equation}\label{16}
W_{\rm kin}= Q_i X^i\;,\qquad\qquad
W^i_{\rm kin}= \f14\;c^2
\varepsilon_{ipq} (\Omega^p - \Omega^p_{\rm prec})\;X^q\;,
\end{equation} where
$Q_i$ characterizes a deviation of the actual world line of
the geocenter from a fiducial world line of a hypothetical spherically-symmetric Earth \cite{kop88}
\begin{equation}\label{17}
Q_i=\partial_i {\bar w}({\bm x}_E)-a_E^i+O(c^{-2})\;.
\end{equation}
Here
$a_E^i={dv^i_E/dt}$ is the
barycentric acceleration of the geocenter. Function
$\Omega^a_{\rm prec}$ describes the relativistic precession
of dynamically non-rotating spatial axes of GCRS with respect to reference quasars
\begin{eqnarray}\label{18}
\Omega_{\rm prec}^i =
\frac{1}{c^2}\,
\varepsilon_{ijk}\,
\left(
-\f32\,v^j_E\,\partial_k {\bar w}({\bm x}_E)
+2\,\partial_k {\bar w}^j({\bm x}_E)
-\f12\,v^j_E\,Q^k
\right).
\end{eqnarray}
The three terms on the right-hand side of this equation
represent the geodetic, Lense-Thirring, and Thomas precessions,
respectively \cite{kop88,iau2}. Dynamic potentials $W_{\rm dyn}$ and $W^i_{\rm dyn}$ are generalizations
of the Newtonian tidal potential in the form of a polynomial starting from the quadratic with respect to ${\bm X}$ terms.

\section{\large IAU SCALING RULES AND THE METRIC TENSOR}

The coordinate transformations between the BCRS and GCRS are found by matching the BCRS and GCRS metric tensors in the vicinity of the world line of the Earth by making use of their tensor properties. The transformations are written as \cite{kop88,iau2}
\begin{eqnarray}
\label{22}
T&=&t - \frac{1}{ c^2} \left[ A + {\bm v}_E\cdot{\bm r}_E \right]
+ \frac{1}{ c^4} \left[ B + B^ir_E^i +
B^{ij}r_E^ir_E^j \right],
\\\label{23}
X^i&=&
r^i_E+\frac 1{c^2}
\left[\frac 12 v_E^i {\bm v}_E\cdot{\bm r}_E + {\bar w}({\bm x}_E) r^i_E
+ r_E^i {\bm a}_E\cdot{\bm r}_E-\frac 12 a_E^i r_E^2
\right],
\end{eqnarray}
where ${\bm r}_E={\bm x}-{\bm x}_E$, functions $A, B, B^i, B^{ij}$ obey equations 
\begin{eqnarray}
\frac{dA}{dt}&=&\f12\,v_E^2+{\bar w}\xe,\label{24}
\\
\frac{dB}{ dt}&=&-\frac{1}{8}\,v_E^4-\frac{3}{ 2}\,v_E^2\,{\bar w}\xe
+4\,v_E^i\,{\bar w}^i+\frac{1}{2}\,{\bar w}^2\xe,\label{25}
\\\label{26}
B^i&=&-\frac{1}{2}\,v_E^2\,v_E^i+4\,{\bar w}^i\xe-3\,v_E^i\,{\bar w}\xe,
\\\label{27}
B^{ij} &=& -v_E^{i} Q_{j}+
2 \p_j {\bar w}^i({\bm x}_E)
-v_E^{i} \p_j {\bar w}\xe+\frac{1}{ 2}
\,\delta^{ij} \dot{{\bar w}}({\bm x}_E)\,,
\end{eqnarray}
where $x_E^i, v_E^i$, and $a_E^i$ are the BCRS position,
velocity and acceleration vectors of the Earth, the overdot stands for the
total derivative with respect to $t$, and one has neglected all terms of the order $O(r^3_E)$.

Earth's orbit in BCRS is almost circular. This makes the right side of equation (\ref{24}) looks like
\begin{equation}\label{u1}
\f12\,v_E^2+{\bar w}\xe=c^2L_C+(\mbox{periodic terms})\;,
\end{equation}
where the constant $L_C$ and the periodic terms have been calculated with a great precision in \cite{irf}. For practical reason the IAU 2000 resolutions recommend to re-scale the BCRS time coordinate $t$ to remove the constant $L_C$ from equation (\ref{u1}). The new time scale is called TDB, and it is defined by equation
\begin{equation}\label{u2}
t_{TDB}=t\left(1-L_B\right)\;,
\end{equation}
where a constant $L_B=L_C+\Delta_C$ is used, instead of $L_C$, in order to take into account the additional linear drift $\Delta_C$ between the GCRS time $T$ and the proper time of clocks on geoid, as explained in \cite{tmsc,irf}.
Time re-scaling changes the Newtonian equations of motion of planets and light. In order to keep the equations of motion invariant entails re-scaling of spatial coordinates and masses of the solar system bodies. These scaling transformations are included to
IAU 2000 resolutions \cite{iau2}. However, the re-scaling of masses, times and spatial coordinates affects the units of their measurement -- the procedure that led to a controversial discussion \cite{ksei}. The change in units can be avoided if one looks at the scaling laws from the point of view of transformation of the metric tensor.

The thing is that the scaling of time and space coordinates can be viewed as a particular choice of the GCRS metric tensor. Indeed, equation (\ref{16}) is a solution of the Laplace equation which is defined up to an arbitrary function of time $Q=Q(t)$ that can be incorporated to
\begin{eqnarray}\label{u3}
W_{\rm kin}&=&Q+Q_i X^i\;,\\
\label{u4}
\frac{dA(t)}{dt}&=&\f12\,v_E^2+{\bar w}\xe-Q\;,
\end{eqnarray}
and, if one chooses $Q=c^2 L_C$, it eliminates the secular drift between times $T$ and $t$ without explicit re-scaling of the time $t$, which is always measured in SI units. It turns out that Blanchet-Damour \cite{bdmass} relativistic definition of mass depends on function $Q$ and is re-scaled automatically in such a way that the Newtonian equations of motion remain invariant \cite{xie}. Introduction of $Q$ to function $W_{\rm kin}$ appropriately transforms the $g_{ij}$ component of the GCRS metric tensor that is formally equivalent to the previously-used re-scaling of the GCRS spatial coordinates. One concludes that introducing the function $Q$ to the GSRC metric tensor without apparent re-scaling of coordinates and masses can be more preferable in updated version of the IAU resolutions as it allows us to keep the SI system of units without changing coordinates and masses.
Similar procedure can be developed for the topocentric metric tensor to take into account the linear drift existing between GCRS time $T$ and the atomic clocks on geoid \cite{tmsc,irf}.

\section{\large PARAMETERIZED COORDINATE TRANSFORMATIONS}

The parameterized post-Newtonian (PPN) formalism \cite{will} is not consistent with the
IAU resolutions.  It limits applicability of the resolutions in testing gravity theories. 
PPN equations of motion depend on two parameters, ${\beta}$ and $\gamma$ \cite{sman} and they
are presently compatible with the IAU resolutions only in the case of $\beta=\gamma=1$.
Rapidly growing precision of astronomical
observations as well as advent of gravitational-wave detectors urgently
demand a PPN theory of relativistic
transformations between the local and global coordinate systems.

PPN parameters ${\beta}$ and $\gamma$ are characteristics of a scalar field which makes the metric tensor different from general relativity.
In order to extend the IAU 2000 resolutions to PPN formalism one used a general class of Brans-Dicke theories \cite{brd}
based on the metric tensor $g_{\alpha\beta}$ and a scalar
field $\phi$ that couples with the metric tensor via
function $\theta(\phi)$. Both $\phi$ and $\theta(\phi)$ are analytic functions
which can be expanded in a Taylor series about their background values $\bar{\phi}$ and $\bar{\theta}$.

The parameterized
theory of relativistic reference
frames in the solar system is built in accordance to the same rules as used in the IAU resolutions. The entire procedure is described in papers \cite{kv,xie}.
The PPN transformations between BCRS and GCRS are found by matching the BCRS and GCRS metric tensors and the scalar field in the vicinity of the world line of the Earth. They have the following form 
\begin{eqnarray}
\label{22q}\hspace{-1cm}
T&=&t - \frac{1}{ c^2} \left[ A + {\bm v}_E\cdot{\bm r}_E \right]
+ \frac{1}{ c^4} \left[ B + B^i\,r_E^i +
B^{ij}\,r_E^i\,r_E^j \right]\,,
\\\label{23q}
X^i&=&
r^i_E+\frac 1{c^2}
\left[\frac 12 v_E^i v_E^jr^j_E +\gamma Qr^i_E+ \gamma{\bar w}({\bm x}_E)r^i_E
+ r_E^i a^j_E r^j_E-\frac 12 a_E^i r_E^2
\right]
\end{eqnarray}
where ${\bm r}_E={\bm x}-{\bm x}_E$, and functions $A(t), B(t), B^i(t), B^{ij}(t)$ obey
\begin{eqnarray}
\frac{dA}{ dt}&=&\frac{1}{2}\,v_E^2+{\bar w}-Q\xe,\label{24q}
\\
\frac{dB}{ dt}&=&-\frac{1}{8}\,v_E^4-\le(\gamma+\frac{1}{ 2}\r)\,v_E^2\,{\bar w}\xe
+2(1+\gamma)\,v_E^i\,{\bar w}^i+\le(\beta-\frac{1}{ 2}\r)\,{{\bar w}}^2\xe,\label{25q}
\\\label{26q}
B^i&=&-\frac{1}{2}\,v_E^2\,v_E^i+2(1+\gamma)\,{\bar w}^i\xe-(1+2\gamma)\,v_E^i\,{\bar w}\xe,
\\\label{27q}
B^{ij} &=& -v_E^{i} Q_{j}+
(1+\gamma) \p_j {\bar w}^i({\bm x}_E)
-\gamma v_E^{i} \p_j {\bar w}\xe+\frac{1}{ 2}
\,\delta^{ij} \dot{{\bar w}}({\bm x}_E)\, .
\end{eqnarray}
These transformations depends explicitly on the PPN parameters ${\beta}$ and $\gamma$ and the scaling function $Q$, and should be compared with those (\ref{22})-(\ref{27}) currently adopted in the IAU resolutions.

PPN parameters ${\beta}$ and $\gamma$ have a fundamental physical meaning in the scalar-tensor theory of gravity along with the universal gravitational constant $G$ and the fundamental speed $c$. It means that if the parameterized transformations (\ref{22q})-(\ref{27q}) are adopted by the IAU, the parameters ${\beta}$ and $\gamma$ are to be considered as new astronomical constants which values have to be determined experimentally.

\section{\large MATCHING IAU RESOLUTIONS WITH COSMOLOGY}

BCRS assumes that the solar system is isolated and space-time is asymptotically flat. This idealization will not work at some level of accuracy of astronomical observations because the cosmological metric has non-zero Riemannian curvature \cite{sfi2}. It may turn out that some, yet unexplained anomalies in the orbital motion of the solar system bodies are indeed associated with the cosmological expansion \cite{p-conf}. Moreover, astronomical observations of cosmic microwave background radiation and other cosmological effects requires clear understanding of how the solar system is embedded to the cosmological model. Therefore, it seems reasonable to incorporate the cosmological metric to the IAU resolutions.

The gravitational field of the solar system has to approximate the cosmological metric tensor at infinity, not a flat metric. The cosmological metric has a number of parameters depending on visible and dark matter and on the dark energy. One considered a universe, driven by a scalar field imitating the dark energy $\phi$, and having a spatial curvature equal to zero \cite{kram,ramk}. The universe is perturbed by a localized distribution of matter of the solar system. The perturbed metric tensor reads
\begin{equation}
\label{c1}
g_{\alpha\beta}=a^2(\eta)f_{\alpha\beta}\;,\qquad f_{\alpha\beta}=\eta_{\alpha\beta}+h_{\alpha\beta}\;,
\ee
where the perturbation $h_{\alpha\beta}$ of the background metric $\bar g_{\a\b}=a^2\eta_{\a\b}$ is caused by matter of the solar system, $a(\eta)$ is a 'radius' of the universe depending on the conformal time $\eta$ related to coordinate time $t$ by simple differential equation $dt=a(\eta)d\eta$. A linear combination of the metric perturbations
\be\l{c1a}
\gamma^{\alpha\beta}=h^{\alpha\beta}-\f12\eta^{\alpha\beta}h\;,
\ee
where $h=\eta^{\alpha\beta}h_{\alpha\beta}$, is more convenient for calculations.

One imposes a cosmological gauge given by \cite{kram,ramk}
\begin{equation}
\label{c2}
\gamma^{\a\b}{}_{|\b}=2H\varphi \d^\a_0\;,
\end{equation}
where a vertical bar denotes a covariant derivative with respect to the background metric $\bar g_{\a\b}$, $\varphi=\phi/a^2$, $H=\dot a/a$ is the Hubble parameter, and the overdot denotes a time derivative with respect to time $\eta$. The gauge (\ref{c2}) generalizes the harmonic gauge of the IAU resolutions for the case of the expanding universe.
The gauge (\ref{c2}) drastically simplifies the field equations. Introducing notations $\gamma_{00}\equiv 4w/c^2$, $\gamma_{0i}\equiv -4w^i/c^3$, and $\gamma_{ij}\equiv 4w^{ij}/c^4$, and splitting Einstein's equations in components, yield
\begin{eqnarray}\label{c5}
\Box\chi-2H\p_\eta \chi+\f52H^2\chi&=&-4\pi G\sigma\;,
\\\label{c6}
\Box w-2H\p_\eta w\phantom{oooo[[::::}&=&-4\pi G\sigma-4H^2\chi\;,\\
\label{c7}
\Box w^{i}-2H\p_\eta w^{i}+H^2 w^{i}&=&-4\pi G\sigma^{i}\;,\\
\label{c8}
\Box w^{ij}-2H\p_\eta w^{ij}\phantom{ooooo}&=&-4\pi G T^{ij}\;,
\end{eqnarray}
where $\partial_\eta\equiv\p/\p\eta$,  $\Box\equiv -c^{-2}\p^2_\eta+{\bm\nabla}^2$, $\chi\equiv w-\varphi/2$, the Hubble parameter $H=\dot a/a=2/\eta$, densities $\sigma=c^{-2}(T^{00}+T^{ss})$, $\sigma^i=c^{-1}T^{0i}$ with $T^{\alpha\beta}$ being the tensor of energy-momentum of matter of the solar system defined with respect to the metric $f_{\alpha\beta}$. These equations extend the equations (\ref{eq1}), (\ref{eq2}) of the IAU resolutions to the case of expanding universe.

Equation (\ref{c5}) describes evolution of the scalar field $\phi$ while equation (\ref{c6}) describes evolution of the scalar perturbation $w$ of the metric tensor. Equation (\ref{c7}) yields evolution of vector perturbations of the metric tensor, and equation (\ref{c8}) describes TT gravitational waves emitted by the solar system. Equations (\ref{c5})--(\ref{c8}) depend on the Hubble parameter and can be solved analytically. The Green functions for these equations have been found in \cite{kram,ramk} and solutions can be smoothly matched with the BCRS metric (\ref{8}), (\ref{9}) of the IAU resolutions.

\textit{This work was promoted by the Research Council Grant FIT-11-020 of the University of Missouri-Columbia. The author is thankful to N. Capitaine and the LOC of Journ\'ees 2010 for travel support.}

\leftskip=5mm


\parindent=-5mm
\smallskip
\begin{thebibliography}{}
\bibitem[1]{ciuf} Ciufolini, I. \& Matzner, R. (eds.)\ 2010, "General Relativity and John Archibald Wheeler", Astrophysics and Space Science Library, 367
\bibitem[2]{leh} L\"ammerzahl, C., Everitt, C.W.F. \& Hehl, F.W. (eds.) 2001, "Gyros, Clocks, Interferometers: Testing Relativistic Gravity in Space", Lecture Notes in Physics, 562
\bibitem[3]{hlnt} Dittus, H., L\"ammerzahl, C., Ni, W.-T. \& Turyshev, S. (eds.) 2008, "Lasers, Clocks and Drag-Free: Technologies for Future Exploration in Space and Tests of Gravity", Springer: Berlin
\bibitem[4]{will} Will, C.M. 1993, "Theory and Experiment in Gravitational Physics", Cambridge University Press: Cambridge
\bibitem[5]{kv} Kopeikin, S. \& Vlasov, I. 2004, "Parametrized post-Newtonian theory of reference frames, multipolar expansions and equations of motion in the N-body problem", \prep, 400, pp. 209-318
\bibitem[6]{xie} Xie, Y., \& Kopeikin, S.\ 2010, "Post-Newtonian Reference Frames for Advanced Theory of the Lunar Motion and a New Generation of Lunar Laser Ranging", Acta Physica Slovaca, 60, pp. 393-495
\bibitem[7]{iau2} Soffel, M., Klioner, S. A., Petit, G., Wolf, P., Kopeikin, S. M., Bretagnon, P., Brumberg, V. A., Capitaine, N., Damour, T., Fukushima, T., Guinot, B., Huang, T.-Y., Lindegren, L., Ma, C., Nordtvedt, K., Ries, J. C., Seidelmann, P. K., Vokrouhlick\'y, D., Will, C. M. \& Xu, C. 2003, "The IAU 2000 Resolutions for Astrometry, Celestial Mechanics, and Metrology in the Relativistic Framework: Explanatory Supplement", \aj, 126, pp. 2687-2706
\bibitem[8]{kopg} Kopeikin, S.M. \& Gwinn, C.R. 2000, "Sub-Microarcsecond Astrometry and New Horizons in Relativistic Gravitational Physics", IAU Colloquium, 180, pp. 303--307
\bibitem[9]{iau06} Capitaine, N., Andrei, A. H., Calabretta, M., Dehant, V., Fukushima, T., Guinot, B., Hohenkerk, C., Kaplan, G., Klioner, S., Kovalevsky, J., Kumkova, I., Ma, C., McCarthy, D. D., Seidelmann, K. \& Wallace, P. T. 2007, "Proposed terminology in fundamental astronomy based on IAU 2000 resolutions", Highlights of Astronomy, 14, pp. 474-475
\bibitem[10]{bk}Brumberg, V.~A. \& Kopejkin, S.~M.\ 1989, "Relativistic reference systems and motion of test bodies in the vicinity of the Earth", Nuovo Cim. B, 103, pp. 63-98
\bibitem[11]{kop88} Kopeikin, S.~M.\ 1988, "Celestial coordinate reference systems in curved space-time", Cel. Mech., 44, pp. 87-115
\bibitem[12]{tmsc} Brumberg, V.~A. \& Kopeikin, S.~M.\ 1990, "Relativistic time scales in the solar system", Cel. Mech. Dyn. Astron., 48, pp. 23-44
\bibitem[13]{irf} Irwin, A.~W. \& Fukushima, T.\ 1999, "A numerical time ephemeris of the Earth", \aap, 348, pp. 642-652    
\bibitem[14]{ksei} Klioner, S., Capitaine, N., Folkner, W., Guinot, B., Huang, T. Y., Kopeikin, S., Petit, G., Pitjeva, E., Seidelmann, P. K. \& Soffel, M. 2009, "Units of Relativistic Time Scales and Associated Quantities", IAU Symposium, 261, pp. 79-84
\bibitem[15]{bdmass} Blanchet, L. \& Damour, T.\ 1989, "Post-Newtonian generation of gravitational waves", Ann. Inst. H. Poincare, Phys. Theor., 50, pp. 377-408
\bibitem[16]{sman} Seidelmann, P. K. 1992, "Explanatory Supplement to the Astronomical Almanac", University Science Books: Mill Valley, California, pp. 281--282
\bibitem[17]{brd} Brans, C.H. \& Dicke, R.H. 1961, "Mach's Principle and a Relativistic Theory of Gravitation", \prd, 124, pp. 925-935
\bibitem[18]{sfi2} Mukhanov, V. 2005, "Physical Foundations of Cosmology", Cambridge University Press: Cambridge   
\bibitem[19]{p-conf} Anderson, J.~D. \& Nieto, M.~M.\ 2010, "Astrometric Solar-System Anomalies", IAU Symposium, 261, pp. 189-197
\bibitem[20]{kram} Kopeikin, S.~M., Ramirez, J., Mashhoon, B. \& Sazhin, M.~V. 2001, "Cosmological perturbations: a new gauge-invariant approach", \pla, 292, pp. 173-180
\bibitem[21]{ramk} Ramirez, J. \& Kopeikin, S. 2002, "A decoupled system of hyperbolic equations for linearized cosmological perturbations", \plb, 532, pp. 1-7
\end{thebibliography}
\end{document}